\documentstyle[12pt]{article}
\begin{document}

\begin{center}
{\large
REPLICA SYMMETRY BREAKING IN AN AXIAL MODEL OF
QUADRUPOLAR GLASS
\\[2mm]}
N. V. Gribova, E.E.Tareyeva
\end{center}

We perform the replica symmetry breaking (RSB) in the vicinity
of the point of instability of the replica symmetric solution
in the model of axial quadrupolar glass. It is shown that the
solution with the first stage RSB is stable against the
second stage RSB. Although there is no reflection symmetry
the 1RSB solution bifurcates continuously from the RS one.
These facts mean that our model can not be associated
with one of two classes usually considered in spin glass
theory.

\newpage

\begin{center}
{\large
REPLICA SYMMETRY BREAKING IN AN AXIAL MODEL OF
QUADRUPOLAR GLASS
\\[2mm]}
N. V. Gribova, E.E.Tareyeva
\end{center}

We perform the replica symmetry breaking (RSB) in the vicinity
of the point of instability of the replica symmetric solution
for the model of axial quadrupolar glass. It is shown that the
solution with the first stage RSB is stable against the
second stage RSB. Although there is no reflection symmetry
the 1RSB solution bifurcates continuously from the RS one.
These facts mean that our model can not be associated
with one of two classes usually considered in spin glass
theory.

\bigskip

1. In recent years the interest to non-Ising spin glasses without
reflection symmetry is renewed. The theory of these models
is often associated with the theory of real structural glasses.
At present time there is no satisfactory microscopic model
of the liquid--glass transition, although a great number of real
and computer experimental data exist, as well as some phemenological
theories. At the same time the investigations of the spin glasses,
orientational glasses and other models with frozen disorder
are actively carried out. They generated lots of ideas and methods
that are used for investigation of other systems, for example neural
networks, proteins, rubbers and real glasses.

Usually two main links between the spin glass theory and the theory
of real glasses are marked. First, the spin glass theory of
some class of models is used as a possible scenarium for the
liquid--glass transition in many-particle systems (see, for example,
\cite{rev1}--\cite{crri}). Second, the methods of the spin glass
theory are used to construct a model for glass-forming systems of
real particles \cite{hnc1}--\cite{parisi00}.

In fact, there is a third link: the approaches are developed
where the transition to the multipole glass state appears as
a part of the liquid--glass transition\cite{chui,tar00}.
It is worth to notice that the physical meaning of the order
parameters is different in different approaches. The problem
considered in this paper can be useful in connection with
the first and the third aspects of links.

As the recent works have shown, experimental characteristics of
relaxation processes are described well enough by the mode coupling
theory \cite{gotze}. The similar equations can be obtained also as
a result of the dynamic investigation of some spin glasses. This
similarity is most pronounced (the first observation of this type
is due to the authors of \cite{kirkp}) in the class of mean-field
spin glasses without reflection symmetry. In this case the static
transition at the temperature $T_c$ (the replica symmetry breaking
-- RSB) is discontinuous and the one-step RSB (1RSB) solution
occurs to be stable, while the full Parisi scheme fails. The
absence of the reflection symmetry means that the internal symmetry
group does not contain the element transforming dynamical variable
${\bf Q}$ to ${\bf -Q}$. This results in cubic terms in
Ginzburg--Landau functional for non-random case \cite{anal} and in
the special form of RSB free energy functional for random
interactions \cite{goel}. The temperature of dynamical transition
$T_d$ in these models is higher than $T_c$. As a result of the
dynamical transition the system is trapped in the state that is
less energetically favorable than the RSB state and stays there
for a long time. The examples of such models are the $p$-spin
model, orientational glasses, random Potts model and etc. They can
be considered to be a prototype of real glasses. Some of these
models, especially the $p$-spin spherical model, have already been
investigated in detail in the middle of the 90ties (see the review
\cite{rev1}). The others are under investigation today
\cite{franzese}--\cite{araijo2}. It is important that the set of
the specific features either appears as a whole or is absent as in
Sherrington-Kirkpatrick model with Parisi solution. The set contains
the absence of the reflection symmetry, the order parameter
discontinuity, the stability of 1RSB solution and the fact that
$T_d > T_c$. So traditionally the idea of two classes of models is
used \cite{rev1,rev2,rev3}.

However, there are some indications that this classification is not
quite correct. For example, the situation is not clear for the
three state Potts model (\cite{tar91}--\cite{gks}).
The 1RSB solution has no discontinuity also in the case of three
spin spherical model for the certain values of the external field
~\cite{last}.

In this paper we consider the quadrupolar glass model with
infinite-range random interaction\cite{tar84,tar86} without
reflection symmetry. We show that in contrast to the usual
classification the 1RSB solution is continuous but stable. May be,
this is due to the influence of the quadrupole operator algebra, that
causes internal fields in the system with the effect similar
to the external field effect in \cite{last}.
(For more details  about internal fields in quadrupolar
glasses see \cite{mosc} and \cite{mosc1}.)

2. So, we investigate the system of N axial quadrupoles that
are in the sites $i,j$ of the regular lattice with the
Hamiltonian \cite{tar84,tar86}

\begin{equation}
H=- \frac{1}{2} \sum_{i \not= j}  J_{ij} Q_i Q_j ,
\label{ham}
\end{equation}
where $Q=3 J^2_{z} - 2$, $J=1, J_z=1, 0,-1$, and
the values of the coupling constants $J_{ij}$
are distributed following the Gaussian law:
$$P(J_{ij}) = ( \sqrt {2 \pi} J)^{-1} \exp
{[-(J_{ij}-J_0)^{2}/2J^2]},$$
and $J_0= \tilde J_0/N$, $J= \tilde J/N^{1/2}$.

Using the standard procedure of the replica method,
we get the expression for the free energy, corresponding
to the Hamiltonian (\ref{ham}) \cite{tar84,tar86}:
$$\frac{\langle F\rangle_J}{NkT} = - \lim_{n \rightarrow 0} \frac{1}{n}
max \left\{ \frac{\tilde J^2n}{(kT)^2} - \frac{1}{2} \sum_{(\alpha \beta)}
( q^{\alpha \beta} )^2 - \frac{1}{2} \sum_\alpha ( x^\alpha)^2
+\right.$$
\begin{equation}
 +\ln Tr \exp \left[ \frac{\tilde J}{kT} \sum_{(\alpha \beta)}
 q^{\alpha \beta} Q^\alpha Q^\beta +\left( \frac{\tilde J_0}{kT}
 + \frac{\tilde J^2}{2(kT)^2}  \right)^{1/2}
\sum_\alpha Q^\alpha x^\alpha -\right.
\label{F1}
\end{equation}
$$-\left. \left.\left( \frac{\tilde J}{kT} \right)^2
\sum_\alpha Q^\alpha \right] \right\}.$$
Here $( \alpha \beta )$ means the sum over the couples of
replicas, $n$ -- the number of replicas, and
$x^\alpha \sim \langle \langle Q^\alpha \rangle_T \rangle_J$,
$q^{\alpha \beta} \sim \langle \langle Q^\alpha Q^\beta \rangle_T
\rangle_J$.
Further, we suppose for simplicity that $J_0=0$.

The case of symmetric replicas was investigated in details in
\cite{tar84,tar86} and the solutions of the equations for the order
parameters were obtained. It was shown that they grow continuously
with the decrease of the temperature. No phase transition
(in the traditional sense) was found. This fact agrees with the experiment
\cite{sul2} for $N_2-Ar$ mixture. There was also obtained
the behaviour of the specific heat with the wide maximum and
linear dependence on the temperature for low temperatures, that
corresponded to the experiment of that time, which had not reached
ultralow temperatures.

Later this model was investigated in detail in many papers
(see, for example \cite{mosc}--\cite{walasek} and references therein),
however the replica symmetry breaking was not carried out.
However the investigation of the stability of the replica symmetric
solution using Almeida-Thouless method \cite{AT} showed the instability
against RSB~\cite{mosc1,walasek}.

In this work we carry out the replica symmetry breaking for the model
(\ref{ham}) in the vicinity of the instability point of the RS
solution. We find the solution in this vicinity, that corresponds
to 1RSB and we show, that it is stable for further replica symmetry
breaking. It is important that despite of absence of reflection
symmetry, 1RSB solution bifurcates continuously that contradicts the
usual classification.

To carry out 1RSB we divide $n$ replicas into $n/m$ groups
with $m$ replicas in each group. We take $q^{\alpha\beta}$
equal to $q_1$ if $\alpha$ and $\beta$ belong to one group
and $q^{\alpha\beta}$ equal to $q_0$ in the opposite case.

Now:
\begin{equation}
\sum_{\alpha \neq \beta} \left( q^{\alpha\beta} \right)^2 =
q_1^2 n(m-1) + q_0^2n(n-m),
\label{qqnrs}
\end{equation}

$$\sum_{\alpha \neq \beta} \left( q^{\alpha\beta} \right)
Q^\alpha Q^\beta= q_0 \left( \sum_1^n Q^\alpha \right)^2 +
 $$

$$\qquad{}+(q_1 - q_0) \left[\left( \sum_{1}^m Q^\alpha \right)^2 +
\ldots+
\left( \sum_{n-m}^n Q^\alpha \right)^2 \right]-$$
\begin{equation}
-2nq_1+q_1 \sum_{1}^n Q^\alpha.
\label{qqQQnrs}
\end{equation}

Substituting (\ref{qqnrs}) and (\ref{qqQQnrs}) into (\ref{F1}),
using the expression for linearization of the exponent:
\begin{equation}
\exp (\lambda a^2) = \frac{1}{\sqrt {2\pi}} \int dx
\exp \left[ - \frac{x^2}{2}+ \sqrt {2\lambda} a x \right]
 \end{equation}
and using new variables
$q_1 \rightarrow t(p+v)$,  $q_0 \rightarrow tp$,  $x \rightarrow
\sqrt {\frac{t^2}{2}}x$,  where  $t = \frac{\tilde J}{kT}$,
we obtain for the free energy:
$$\frac{F}{NkT} = -t^2 + \frac{t^2 x^2}{4}
+ \frac{t^2}{4} \left( -m p^2 + (p+v)^2 (m-1) + 4 (p+v) \right)-$$
\begin{equation}
- \frac{1}{m} \int dy^G \ln \int dz^G \Psi^m \left( \theta
\right), \label{F2} \end{equation}
where
$$ da^G = \frac{1}{\sqrt
{2\pi}} da e^{-a^2/2}, $$
$$ \Psi = 2 e^\theta + e^{-2 \theta},$$
$$\theta = t y \sqrt {p} + tz\sqrt{v} + \frac{t^2}{2} ( p+v-2+x ),$$
and $x,p,v,m$ satisfy the equations that express the extremum
condition  of the functional ~(\ref{F2}):
\begin{equation}
x = \int dy^G \frac{\int dz^G \Psi^{m-1} \Psi'}{ \int dz^G
\Psi^m},
\label{x}
\end{equation}

\begin{equation}
p + v = \int dy^G \frac{\int dz^G \Psi^{m-2} \left( \Psi' \right)^2}
{ \int dz^G \Psi^m},
\label{pv}
\end{equation}

\begin{equation}
p = \int dy^G \left[ \frac{\int dz^G \Psi^{m-1} \Psi'}{ \int dz^G \Psi^m}
\right]^2,
\label{p}
\end{equation}

$$-\frac{t^2}{4} m \left( (p+v)^2 -p^2 \right) =
\frac{1}{m} \int dy^G \ln \int dz^G \Psi^m -$$
\begin{equation}
-\int dy^G \frac{\int dz^G \Psi^m \ln \Psi }{ \int dz^G \Psi^m}.
\label{m}
\end{equation}

From the expression for free energy (\ref{F2}) we get the entopy
in form:
$$\frac{S}{Nk}= -t^2 -\frac{3}{4}t^2 x^2 - \frac{3}{4}t^2\left[
-m p^2 +(p+v)^2(m-1)\right] -t^2(p+v)+2t^2x+$$
$$+\frac{1}{m} \int dy^G \ln \int dz^G \Psi^m \left( \theta
\right).$$

3. Expanding (\ref{x}--\ref{m}) in $t$ at $t \rightarrow 0$,
it is easy to show, that there is only one solution for the
system of new equations at high temperatures and it is the same
as for RS equations \cite{tar84,tar86}. One can obtain the RS
equations from (\ref{F2}--\ref{m}) if $ v=0 $:

\begin{equation}
x' =2 \int dy^G
 \left[ \frac{e^{\theta'}-e^{-2 \theta'}}{2 e^{\theta'}+
 e^{-2 \theta'}} \right],
\label{x_al}
\end{equation}

\begin{equation}
p' =4 \int dy^G
 \left[ \frac{e^{\theta'}-e^{-2 \theta'}}{2 e^{\theta'}+e^{-2 \theta'}}
 \right]^2,
\label{p_al}
\end{equation}

$$\theta' = t y \sqrt {p'} + \frac{t^2}{2} ( p'+x'-2 ).$$

The solutions $x = x'$, $p = p'$, $v = 0$ are the solutions
of the system of equations (\ref{x})--(\ref{m}) for any $m$
at all temperatures.
Now let us investigate the behaviour of 1RSB system near
the possible point of instability. To do this we expand
the free energy~(\ref{F2}) in series near the arbitrary
point $t_c$:
\begin{equation}
\begin{array}{l} F = F_0 + \frac{1}{2} x_1^2 \left( F_{xx} + \tau
F_{xxt} \right) + \frac{1}{2}p_1^2 \left( F_{pp} + \tau F_{ppt}
 \right) + \\ \qquad{}  + x_1 p_1 \left( F_{xp} + \tau F_{xpt}
  \right) + \frac{1}{2}v^2 \left( F_{vv} + \tau F_{vvt} \right) +
\\ \qquad{}+ x_1 v \left( F_{vx} + \tau F_{vxt} \right) + p_1 v
\left( F_{vp} + \tau F_{vpt} \right) +  \\ \qquad{}
 +\frac{1}{6} x_1^3 F_{xxx} +\frac{1}{6}p_1^3 F_{ppp} +
 \frac{1}{6} v^3 F_{vvv} +\\
 \qquad{} +\frac{1}{2}x_1^2\left( p_1 F_{xxp} + v F_{xxv} \right) +
  + \frac{1}{2}x_1^2\left( p_1 F_{xxp} + v F_{xxv} \right)+\\
  \qquad{} +\frac{1}{2}p_1^2\left( x_1 F_{ppx} + v F_{ppv} \right) +
 +\frac{1}{2}v^2
\left( x_1 F_{vvx} + p_1 F_{vvp} \right) +\\
\qquad{}+ x_1 p_1 v F_{xpv},
\end{array}
\label{Frazl}
\end{equation}
where $F_0$ -- free energy of the RS solution, $\tau=t-t_c$,
 and $x_1$, $p_1$, $v$ are deviations from RS solutions
($p = p' + p_1,x = x' + x_1$). Let us look for small solutions
of the equations for the extremum condition of the functional
~(\ref{Frazl}), following the general bifurcation theory
(see, for example ~\cite{VaTr}). Introducing new variable
$A = F/(Nkt)$  and using the expressions for $A_{ab}$ from Appendix 1,
we write the extremum condition in variables $x$, $p$, $v$ in the form:

\begin{equation}
x_1 A_{xx} + p_1 A_{px} - v (m-1) A_{px} = B_x,
\label{Frazl1}
\end{equation}

\begin{equation}
x_1 A_{px} + p_1 A_{pp} - v (m-1) A_{pp} = B_p,
\label{Frazl2}
\end{equation}

\begin{equation}
x_1 A_{px} + p_1 A_{pp} + v [A_{pp} - 2 m D] =B_v,
\label{Frazl3}
\end{equation}
where $D$ is defined by the expression
$$ A_{vv} = - (m - 1) [ A_{pp} - 2 m D],$$
$B_x$, $B_p$, $B_v$ are bilinear in $x_1$, $p_1$, $v$, $\tau$.
The equations (\ref{Frazl1} -- \ref{Frazl3}) have always the trivial
solution that corresponds to RS solution. And at the singular
point which is determined from the condition that the determinant
of the linear homogeneous system is equal to zero:
\begin{equation}
det \left(
\begin{array}{ccc}
  A_{xx}& A_{xp}& A_{xv} \\
  A_{px}& A_{pp}& A_{pv} \\
  A_{vx}& A_{vp}& A_{vv}
\end{array}
\right) = 0
\label{A}
\end{equation}
the solution ceases to be unique. The condition ~(\ref{A})
can be rewritten in the form:
$$ m \left( m - 1 \right) \left( A_{pp} - 2 D \right)
\left( A_{xx} A_{pp} - A_{px}^2 \right)  = 0. $$

Here the coefficients are for RS solution.
Using ~\cite{tar84,tar86}, it is easy to see that
$$D' = A_{xx} A_{pp} - A_{px}^{2}$$
nowhere turns to zero (see also ~\cite{walasek}), and
the condition
\begin{equation}
A_{pp} - 2 D = 0
\label{det}
\end{equation}
defines the temperature of the possible bifurcation:
$$t = t_c = \frac{1}{1,367}.$$
At this point RS solution is
\begin{equation}
x'(t_c) = -0,581; ~~~~~~
p'(t_c) = 1,449.
\label{cyfry}
\end{equation}
If the condition ~(\ref{det}) is fulfilled, then the fourth
equation of the system -- the extremum condition in
$m$ -- has the form:
\begin{equation}
x_1 A_{px} + p_1 A_{pp} - v (m-1) A_{pp} = B_m,
\label{Frazl4}
\end{equation}
so in fact it plays a role of the compatibily condition.

4. To find small solutions of the system
(\ref{Frazl1} -- \ref{Frazl3}, ~\ref{Frazl4})
in the vicinity of $t_c$, we have to (see ~\cite{VaTr}) solve the system
of two equations with the determinant not equal to zero $D'\neq 0$
for variables $x_1$ and $p_1$, and to define $v$ and $m$
from the bifurcation equations. From the system
\begin{equation}
x_1 A_{xx} + p_1 A_{px} = v (m-1) A_{px} + B_x,
\label{Frazl1a}
\end{equation}

\begin{equation}
x_1 A_{px} + p_1 A_{pp} = v (m-1) A_{pp} + B_p
\label{Frazl2a}
\end{equation}
we obtain
\begin{equation}
x_1 = \frac{B_p A_{px} - B_x A_{pp}}{D'},
\label{xfull}
\end{equation}
\begin{equation}
p_1 = v (m - 1) + \frac{B_x A_{px} - B_p A_{xx}}{D'},
\label{pfull}
\end{equation}
so in the main order
\begin{equation}
x_1 = 0; ~~~~~~
p_1 = v (m-1).
\label{prib}
\end{equation}

Substituting the obtained solutions into the initial equations
(\ref{Frazl1} - \ref{Frazl3}, ~\ref{Frazl4})
(in the l.h.s. we need to substitute the full solutions
~(\ref{xfull}) and ~(\ref{pfull}), while for r.h.s. ~(\ref{prib})
is enough), we obtain the bifurcation equations, that in our case
lead simply to
\begin{equation}
B_p = B_v;~~~~~~~  B_m = B_v
\label{}
\end{equation}
or
\begin{equation}
\tau G_1 = - v \frac{t^6}{16} (G_2 + m G_3) ;
\label{tauv1}
\end{equation}
\begin{equation}
\tau \frac{2 m - 1}{m} G_1 = - v \frac{t^6}{48} [- G_2 + 2 m
(G_2 - G_3) + 3 m^2 G_3].
\label{tauv2}
\end{equation}
The expressions for the coefficients $G_i$ and their numerical
values for RS solution are given in the Appendix 1.
Solving ~(\ref{tauv1}) -- ~(\ref{tauv2}), we finally obtain:
\begin{equation}
m = \frac{G_2}{2 G_2 + G_3} = 0,427,
\label{mnum}
\end{equation}
\begin{equation}
v = \tilde v \tau;~~~~ \tilde v = - \frac{16}{t_{c}^{6}}
\frac{G_1}{G_2 + m G_3} = 3,13
\label{vnum}
\end{equation}
So, we managed to find the small 1RSB solution of the system
~(\ref{x}--\ref{m}) near $T_c$. The 1RSB  solution
$x = x', p = p' + \tilde v \tau (m - 1),
v = \tilde v \tau$ at $m = 0,427$ bifurcates continuously
from the RS solution.

5. Now let us investigate the stability of 1RSB against the
further replica symmetry breaking (2RSB). Now we divide each
former group of replicas with $m$ elements into $m/m_1$
groups with $m_1$ elements. Parameter $q_{\alpha \beta }$
is assigned $p_3$, if replicas $\alpha $ and $\beta $
belong to the same smallest group (the number of $p_3$ is $\frac{n
(m-1)}{2}$), $p_2$ -- if $\alpha $ and $\beta $ belong to the
same "middle" group, but to the different smallest groups
(they are $\frac{n (m-m_1)}{2}$ there), and $p_0$ -- if
$\alpha $ and $\beta $ belong to the different groups of $m$
replicas (their number is $\frac{n(n-m)}{2}$). As usual
(see, for example, ~\cite{som}), we deal with fluctuations
inside one group of $m$ replicas and intergroup fluctuations,
connected with the corresponding overlap of replicas
$q_{\alpha \beta}$. The term in the free energy, corresponding
to the intergroup replicon mode, which defines the stability of
1RSB solution in our case, has the form:
\begin{equation}
\Delta F_1 =  \frac{1}{2}v_{1}^{2} (m - 1) (m - m_1) (m_1 - 1)
(P_0 - 2 Q_4 + R_5),
\label{del1}
\end{equation}
and
$$P_0 = 2 - 2 t^2 [4 - 4 x + p + v - (p + v)^2],$$
$$Q_4 = - 2 t^2 [2 (p + v) - (p + v)^2 - t_3],$$
$$R_5 = - 2 t^2 [r_4 - (p + v)^2],$$
where $v_1$ ¨ $w_1$ (see Appendix 2) are defined by the expressions
$$p_3 = w_1 - (m - m_1) v_1,$$
$$p_2 = w_1 + (m_1 - 1) v_1.$$
The eigenvalue is
$$\Lambda ' = P_0 - 2 Q_4 + R_5 = 2 - 2 t^2 [4 - 4 x - 3 (p+v)
+2 t_3 + r_4],$$
where
\begin{equation}
t_{3} = \int dy^G \frac{ \int dz^G \Psi^{m-3} \left( \Psi'
\right)^3} { \int dz^G \Psi^m},
\label{t3}
\end{equation}
\begin{equation}
r_{4} = \int dy^G \frac{\int dz^G \Psi^{m-4} \left( \Psi'
\right)^4} { \int dz^G \Psi^m}.
\label{r4}
\end{equation}

The other terms in the free energy are given in the
Appendix 2 ~(\ref{del2f}).
With the help of RS and 1RSB solutions we can obtain the value
of $\Lambda'$ near $t_c$. Taking into account that
$\Lambda '(t_c) = 0$ and expanding $t_3$ and $r_4$ in series
in variables $p$, $v$ and $t$, using the expressions
\begin{equation}
\frac{\partial t_3}{\partial p} = 3 t^2 \langle \alpha W (W +
\alpha ^2 )\rangle,
\label{t3p}
\end{equation}

\begin{equation}
\frac{\partial t_3}{\partial v} = \frac{\partial t_3}
{\partial p} - 3 m t^2 \langle \alpha ^3 W  \rangle,
\label{t3v}
\end{equation}

\begin{equation}
\frac{\partial r_4}{\partial p} = 2 t^2 \langle \alpha^2 W (3W +
2 \alpha ^2) \rangle,
\label{r4p}
\end{equation}

\begin{equation}
\frac{\partial r_4}{\partial v} = \frac{\partial v_4}
{\partial p} - 4 m t^2 \langle \alpha ^4 W  \rangle,
\label{r4v}
\end{equation}
we obtain for the solution ~(\ref{prib}), ~(\ref{mnum}),
~(\ref{vnum}):
$$\Lambda ' = 6t_c^2 m \tau \tilde
v (1 - 2 t_c^2 G_4).$$
Here we take $G_4$ for RS  solution:
$$G_4 = \langle W (\alpha ^4 + 2 \alpha ^3 - \alpha ^2 -
2 \alpha ) \rangle = 0,27 ,$$
so finally
$$\Lambda ' = 2,97 \tau, $$
that means the stability of 1RSB solution in this
vicinity at $T<T_c$ and $m_1<m<1$.

6. In conclusion, we have investigated the behaviour of the model
(\ref{ham})
near the instability point $t_c$ of its RS solution. We have shown
that in the vicinity of $t_c$ there is a more favorable solution
which
corresponds to the one-step replica symmetry breaking. We have
also shown, that the new  solution is stable relative to further
replica symmetry breaking, i.e. the full Parisi scheme does not
work in this case, at least in its classical variant.
It is worth to notice that in spite of absence of reflection
symmetry,
1RSB solution bifurcates continuously from the RS solution,
that contradicts the idea of division in two classes.

However, one can await that at further decrease  of the
temperature,  our 1RSB solution will become unstable.
For complete investigation of this case it is necessary to
solve the system of equations (\ref{x}--\ref{m}) for all
temperatures and use  the obtained solution for the definition
of the sign of $\Lambda_{1'}$ everywhere. May be, the
low-temperature form of the 1RSB solution will explain the
behaviour of experimental data \cite{mosc15} at ultralow
temperatures $ T \approx 0,05\, K $, where the deviations
from the RS solution are noticed, i.e. the linear dependence
of the specific heat on $T$ changes to the quadratic one.

Authors thank V.N. Ryzhov and T.I. Shchelkacheva for
helpful discussions and valuable comments.

This work was supported in part by Russian Foundation for
Basic Researches (Grant No. 02-02-16621).

\section{Appendix 1}
Let us take into account that $v=0$ for the RS solution. So
$\Psi$ does not depend on $z$, and the internal integrals in
(\ref{F2} -- ~\ref{p}) can be taken directly. Let us denote:
$$\left.{\frac{\Psi'}{\Psi}}\right|_{t=t_c,\, x=x_0,\, p=p_0,\, v=0} =
\alpha,$$
$$ W = \alpha^2 + \alpha - 2,$$
$$ \langle . . . \rangle = \int dy^G ...\quad.$$

In these notations
$$A_{xx} = 1 + \frac{t^2}{2} \langle W \rangle,$$
$$A_{px} = - \frac{t^2}{2} \langle W \alpha \rangle,$$
$$A_{pp} = - 1 + t^2 \langle W (W + 2 \alpha^2) \rangle,$$
$$A_{vx} = - (m - 1) A_{px},$$
$$A_{pv} = - (m - 1) A_{pp},$$
$$A_{vv} = - (m - 1) [ A_{pp} - 2 m D],$$
$$D = t^2 \langle W \alpha^2 \rangle,$$
$$A_{pp} - 2 D = -1 + t^2 \langle W^2 \rangle,$$

At the bifurcation point $t_c = \frac{1}{1,367}$,
$x' = - 0,581$, $p' = 1,449$ and we have for the RS solution
$$\langle W \rangle = - 1,132,$$
$$\langle W \alpha \rangle = 0,6035,$$
$$\langle W \alpha^2 \rangle = - 0,997,$$
$$\langle W \alpha^3 \rangle = 1,232,$$
$$\langle W \alpha^4 \rangle = - 1,981.$$
\medskip

The coefficients $B_x, B_p, B_v$ contain the third
and $B_m$ -- the fourth derivatives of $A$ at the
point $t_c$, and their numerical values are obtained
from the RS solution. With their help we get:

$$G_1 = - t + t^5 \langle W [(x'(t_c) - 2) (2 \alpha^3 +
3 \alpha ^2 - 3 \alpha  - 2) +$$ $$+ p'(t_c) (- 10 \alpha ^4 +
18 \alpha ^3 + 15 \alpha ^2 + 19 \alpha - 6)] \rangle = -0,4,$$
$$G_2 = 4 \langle W [4 \alpha^4 + 8 \alpha ^3 - 3 \alpha ^2 -
7 \alpha - 2] \rangle = 11,8,$$
$$G_3 = 4 \langle W [- 10 \alpha^4 - 20 \alpha ^3 + 12 \alpha ^2
+ 22 \alpha + 10] \rangle = 4.$$

\section{Appendix 2}
\begin{equation}
\begin{array}{l}
4 \Delta F_2  =   p_0^2 m [ - 2 P_1 + 4 m Q_1
- 4 (m-1) Q_3 - 6 m^2 R_1 + 8 m (m-1) R_3 - 2 (m-1)^2 R_7] + \\
\qquad{} + w_1^2 (m-1) [2 P_0 + 4 Q_4 (m-2) + R_5 (m-2)(m-3) -
R_6 m (m-1)] +\\ \qquad{} + p_0 w_1 4 m (m-1) [ -2Q_2 + m R_2 -
(m-2) R_4].
\label{del2f}
\end{array}
\end{equation}
Here the coefficients $P$, $Q$, $R$ are for 1RSB solution
and are as follows:
$$P_1 = 2 - 2 t^2 [4 - 4 x + p - p^2],$$ $$Q_1 =
- 2 t^2 [2 p - p^2 - t_{111}],$$ $$Q_2 = - 2 t^2 [2 p - p (p +
v) - t_{21}],$$ $$Q_3 = - 2 t^2 [2 (p + v) - p^2 - t_{21}],$$
$$R_1 = - 2 t^2 [r_{1111} - p^2],$$ $$R_2 = - 2 t^2 [r_{211} - p
(p + v)],$$ $$R_3 = - 2 t^2 [r_{211} - p^2],$$ $$R_4 = - 2 t^2
[r_{31} - p (p + v)],$$ $$R_6 = - 2 t^2 [r_{22} - (p + v)^2],$$
$$R_7 = - 2 t^2 [r_{22} - p^2],$$
where
\begin{equation}
t_{111} = \int dy^G \left[ \frac{\int dz^G \Psi^{m-1} \Psi'}{
\int dz^G \Psi^m} \right]^3,
\label{t111}
\end{equation}
\begin{equation}
t_{21} = \int dy^G  \frac{\int dz^G \Psi^{m-1} \Psi'
\int dz^G \Psi^{m-2} \left (\Psi'\right)^2}{\left[
\int dz^G \Psi^m\right]^2},
\label{t21}
\end{equation}
\begin{equation}
r_{1111} = \int dy^G \left[ \frac{\int dz^G \Psi^{m-1} \Psi'}{
\int dz^G \Psi^m} \right]^4,
\label{r11}
\end{equation}
\begin{equation}
r_{31} = \int dy^G \frac{\int dz^G \Psi^{m-1} \Psi'
\int dz^G \Psi^{m-3} \left( \Psi'
\right)^3}
{ \left[\int dz^G \Psi^m  \right]^2},
\label{r31}
\end{equation}
\begin{equation}
r_{22} = \int dy^G \left[ \frac{ \int dz^G \Psi^{m-2}
\left (\Psi'\right)^2}{ \int dz^G \Psi^m} \right]^2.
\label{r22}
\end{equation}

\bigskip

\end{document}